\documentclass[11pt,a4paper]{article}

\usepackage{latexsym,amsmath,amsthm,amssymb}
\usepackage{epsfig,graphics}
\usepackage{color}
\usepackage[ruled]{algorithm2e}
\usepackage{fullpage}
\usepackage{pstricks,pst-node}
\usepackage{subfig}
\usepackage{paralist}

\newtheorem{theorem}{Theorem}
\newtheorem{definition}{Definition}

\newtheorem{lemma}{Lemma}

\newcommand{\ceil}[1]{\left\lceil {#1} \right\rceil}

\newcommand{\calC}{\mathcal{C}}
\newcommand{\calD}{\mathcal{D}}

\newcommand{\calP}{\mathcal{P}}
\newcommand{\calQ}{\mathcal{Q}}
\newcommand{\calR}{\mathcal{R}}

\newcommand{\calT}{\mathcal{T}}

\newcommand{\calX}{\mathcal{X}}

\newcommand{\vc}{\textsc{Minimum Vertex Cover}}
\newcommand{\rvc}{\textsc{Minimum Rectangle Vertex Cover}}
\newcommand{\is}{\textsc{Maximum Independent Set}}
\newcommand{\ris}{\textsc{Maximum Rectangle Independent Set}}

\newcommand{\eps}{\varepsilon}

\newcommand{\opt}{\textsc{opt}}

\newcommand{\poly}{\text{\normalfont poly}}
\newcommand{\tw}{\text{\normalfont tw}}

\begin{document}

\title{Minimum Vertex Cover in Rectangle Graphs}

\author{%
Reuven Bar-Yehuda%
\thanks{Department of Computer Science, Technion IIT, Haifa 32000, Israel.
{\tt reuven@cs.technion.ac.il}.}
\and
Danny Hermelin%
\thanks{Max-Planck-Institut f\"{u}r Informatik, Saarbr\"{u}cken, Germany. 
{\tt hermelin@mpi-inf.mpg.de}.}
\and
Dror Rawitz%
\thanks{School of Electrical Engineering, Tel-Aviv University,
Tel-Aviv 69978, Israel. {\tt rawitz@eng.tau.ac.il}.}
}


\maketitle

\begin{abstract}
We consider the \vc\ problem in intersection graphs of axis-parallel
rectangles on the plane.  We present two algorithms: The first is an
EPTAS for non-crossing rectangle families, rectangle families $\calR$
where $R_1 \setminus R_2$ is connected for every pair of rectangles
$R_1,R_2 \in \calR$.
This algorithm extends to intersection graphs of pseudo-disks.
The second algorithm achieves a factor of $(1.5 + \varepsilon)$ in
general rectangle families, for any fixed $\varepsilon > 0$, and works
also for the weighted variant of the problem. Both algorithms exploit
the plane properties of axis-parallel rectangles in a non-trivial way.
\end{abstract}




\section{Introduction}
\label{Section: Introduction}

In this paper we are concerned with the \rvc\ problem: Given a set
$\calR=\{R_1,\ldots,R_n\}$ of (weighted) axis-parallel rectangles in
the plane, find a minimum size (weight) subset of rectangles in
$\calR$ whose removal leaves the remaining rectangles in $\calR$
pairwise disjoint, \emph{i.e.} no pair of remaining rectangles share a
common point.  This problem is a special case of the classical \vc\
problem, which asks to find a minimum weight subset of vertices in a
given graph, whose removal leaves the graph without edges.  When the
input graph is a \emph{rectangle graph}, an intersection graph of axis
parallel rectangles in the plane, and the rectangle representation of
the input graph is given alongside the input,
\vc\ becomes $\rvc$.

\vc\ is one of the most extensively studied combinatorial
problems in computer science, a study dating back to K\"{o}nig's
classical early 1930s result~\cite{Konig1931}, and probably even prior
to that. Karp proved that the problem is NP-complete in his famous
list of fundamental NP-complete problems~\cite{Karp1972}, while Garey
and Johnson extensively used \vc\ as an intermediate problem in many
of their early NP-completeness reductions~\cite{GareyJohnson1979}.
Since then, \vc\ played a pivoting role in the development of both
approximation algorithms~\cite{Hochbaum1997,Vazirani2003}, and the
theory of parameterized
complexity~\cite{DowneyFellows1999,FlumGrohe2006,Niedermeier2006}, the
two main disciplines for coping with the widespread phenomena of
NP-hardness.

From the perspective of approximation algorithms, \vc\ has many
polynomial-time algorithms achieving an approximation ratio of
2~\cite{Bar-YehudaEven1981,Clark1983,Hochbaum1982,NemhauserTrotter1975},
the first of these given in Nemhauser and Trotter's fundamental
paper~\cite{NemhauserTrotter1975}.  Moreover, the problem is known to
be approximable in within $2-\frac{\lg\lg n}{2\lg
n}$~\cite{Bar-YehudaEven1985,MonienSpeckenmeyer1985}, within
$2-\frac{\ln\ln n}{\ln n}(1-o(1))$~\cite{Halperin2002} and even within
$2-\Theta(\frac{1}{\sqrt{\log n}})$~\cite{Karakostas05}.  On the other
hand, it is also known that \vc\ is inapproximable within a factor of
$10\sqrt{5}-21 \approx 1.36$, unless P$=$NP~\cite{DinurSafra2002}.
There are however many natural special-case graph classes for which
one can improve on this barrier. For instance, in the class of
interval graphs, which can be thought of as one dimensional analogs of
rectangle graphs, \vc\ is polynomial-time solvable~\cite{Gavril1972}.
In planar graphs, the problem is known to admit a polynomial-time
approximation scheme
(PTAS)~\cite{LiptonTarjan1980,Bar-YehudaEven1982,CNS81}, and even an
efficient PTAS (EPTAS) due to Baker's seminal framework for NP-hard
planar graph problems~\cite{Baker1994}.

The dual problem of \rvc\ is the \ris\ problem: Given a family of
axis-parallel rectangles in the plane, find a maximum size (or weight)
subset of pairwise disjoint rectangles.  This problem has been
extensively studied in the computational geometry community, and has
several applications in data
mining~\cite{ChalermsookChuzhoy2009,KhannaMuthukrishnanPaterson1998},
automated label
placement~\cite{AgarwalVanKreveldSuri1998,DoerschlerFreeman1992,Freeman1991},
and in network resource allocation with advance reservation for line
topologies~\cite{ChalermsookChuzhoy2009,Lewin-EytanNaorOrda2002},
which also apply to \vc\ in rectangle graphs (see
Section~\ref{Section: Applications and Motivation}).  Fowler \emph{et
al.}~\cite{FowlerPatersonTanimoto1981} showed that \is\ in rectangle
graphs is NP-complete, implying the NP-completeness of \vc\ in
rectangle graphs.  Asano~\cite{Asano1991} showed that \is\ and \vc\
remain NP-hard even in intersection graphs of unit squares. There have
been several $O(\lg n)$ approximation algorithms independently
suggested for this
problem~\cite{AgarwalVanKreveldSuri1998,Bermanetal2001,Chan2004,KhannaMuthukrishnanPaterson1998}.
Lewin-Eitan \emph{et al.}~\cite{Lewin-EytanNaorOrda2002} devised a
$4q$-approximation algorithm for the problem, where $q$ is the size of
the maximum clique in the input graph.  Recently, Chalermsook and
Chuzhoy~\cite{ChalermsookChuzhoy2009} were able to break the $\lg n$
approximation barrier by devising a sophisticated $O(\lg \lg n)$
randomized approximation algorithm.  A simpler $O(\lg n/\lg\lg
n)$-approximation algorithm was given in~\cite{ChanHar-Peled2009}.
There are also many special cases in which \ris\ admits a
polynomial-approximation scheme
(PTAS)~\cite{AgarwalVanKreveldSuri1998,Bermanetal2001,Chan2003,ErlebachJansenSeidel2005}.

In contrast to the vast amount of research devoted for \ris\, there
has been surprisingly very little focus on the \rvc\ problem.
Nevertheless, some of the results for \ris\ carry through to \rvc.
For instance, the result of Fowler \emph{et
al.}~\cite{FowlerPatersonTanimoto1981} implies that \rvc\ is
NP-hard. Also, by applying the Nemhauser and Trotter Theorem (see
Section~\ref{Section: Preliminaries}) as a preprocessing step, any
PTAS for \ris\ can be converted into a PTAS for
\rvc.  Thus, the results in~\cite{AgarwalVanKreveldSuri1998}
imply that \rvc\ has a PTAS when all rectangles have equal height,
while~\cite{Chan2003} gives a PTAS when all rectangles are squares.
Erlebach \emph{et al.}~\cite{ErlebachJansenSeidel2005} gave an
explicit PTAS for \rvc\ in bounded aspect-ratio rectangle families
without using the Nemhauser and Trotter procedure.  Finally, we
mention the work by Chan and Har-Peled~\cite{ChanHar-Peled2009} who
devised a PTAS for \is\ in families of \emph{pseudo-disks}, which are
families of regions on the plane such that the boundaries of every
pair of regions intersect at most twice.
This result implies a PTAS for \rvc\ in non-crossing rectangle
families.


\subsection{Related Work}
\label{Section: Related Work}

\vc\, and its dual counterpart \is\, have been previously
studied in many geometric intersection graphs other than rectangle
graphs.  Gavril~\cite{Gavril1972} gave a polynomial-time algorithm for
both of these problem in chordal graphs, intersection graphs of
subgraphs of a tree. Apostolico \emph{et
al.}~\cite{ApostolicoAtallahHambrusch1992} gave a polynomial-time
algorithm for these two problems in intersection graphs of chords on a
circle, which were later improved by Cenek and
Stuart~\cite{CenekStuart2003}, while Golumbic and
Hammer~\cite{GolumbicHammer1988} gave a polynomial-time algorithm for
intersection graphs of arcs on a circle which was later improved
in~\cite{HsuTsai1991}. A good survey of many generalizations of these
results can be found in~\cite{Golumbic1980,GolumbicTrenk1985}.
Hochbaum and Maass and later Chleb\'{\i}k and Chleb\'{i}kov\'{a}
considered intersection graphs of $d$-dimensional boxes in
$\mathbb{R}^d$~\cite{ChlebikChlebikova2005,HochbaumMaass1985}, while
Erlebach \emph{et al.}~\cite{ErlebachJansenSeidel2005} considered
intersection graphs of general fat objects in the plane.
In~\cite{Bar-Yehudaetal2002,ButmanHermelinLewensteinRawitz2007},
approximation algorithms were suggested for \is\ and \vc\ in the class
of multiple-interval graphs.




\subsection{Applications and Motivation}
\label{Section: Applications and Motivation}

Automated label placement is a central problem in geographic
information systems which has been extensively studied in various
settings~\cite{AgarwalVanKreveldSuri1998,DoerschlerFreeman1992,Freeman1991}.
The basic problem is to place labels around points in a geographic
maps, where the labels are often assumed to be
rectangles~\cite{AgarwalVanKreveldSuri1998} which are allowed to be
positioned at specific places adjacent to their corresponding points
in the map. The usual criterion for a legal placement is that all
rectangles are pairwise disjoint.  Subject to this constraint, a
natural optimization criteria is to minimize the number of labels to
be removed so as the remaining labels form a legal placement. This is
exactly the \rvc\ problem.


\rvc\ can also be used to model shared-resource scheduling scenarios 
where the requests are given in advance to the system.  Consider the
typical critical-section scheduling problem occurring in all modern
operating system: A set of programs request access to a shared
resource in memory for read$\backslash$write purposes. The goal of the
operating system is to serve as many requests as possible, so long as
no two programs access the same memory entries simultaneously, to
avoid obvious data-consistency hazards. In a simplified variant of
this problem, one can assume that all programs have a single request
to fixed array of registers in memory, and this request occurs during
a fixed interval of their running time. If these requests are known
beforehand to the operating system, the problem of minimizing the
number of programs not to be served can naturally be modeled as
\rvc\ by using the $x$-axis to measure the shared memory array, and
the $y$-axis to measure program execution-time.



\subsection{Results and Techniques}
\label{Section: Results and Techniques}

In this paper we present two approximation algorithms for the \rvc\
problem. For a pair of rectangles $R_1$ and $R_2$ in our input set of
rectangles $\calR$, we say that $R_1$ and $R_2$ \emph{cross} if they
intersect, but neither rectangle contains a corner of the other
rectangle.  This is equivalent to requiring that $R_1 \setminus R_2$
is connected for every $R_1, R_2 \in \calR$.
(We assume w.l.o.g.\ that the rectangles are in general position.)
We say that $\calR$ is \emph{non-crossing} if there is no pair of
crossing rectangles in $\calR$.  Our first algorithm is an EPTAS for
\rvc\ in non-crossing rectangle families:

\begin{theorem}
\label{Theorem: Non-Crossing Rectangles}%
Given any $\eps >0$, \rvc\ in non-crossing rectangle families
can be approximated within $(1+\eps)$ in $2^{\poly(1/\eps)}
\cdot \poly(n)$ time.
\end{theorem}

We mention that \rvc\ in non-crossing rectangle families is
NP-hard according to~\cite{Asano1991}. Theorem~\ref{Theorem:
Non-Crossing Rectangles} generalizes the PTAS result of Agarwal
\emph{et al.}~\cite{AgarwalVanKreveldSuri1998} and
Chan~\cite{Chan2003} for squares and equal height rectangles,
and it also handles several families of rectangles which cannot
be handled by the PTAS of Erlebach \emph{et
al.}~\cite{ErlebachJansenSeidel2005}. In terms of time
complexity, our algorithm dramatically improves on all these
algorithms, and also on the algorithm of Chan and
Har-Peled~\cite{ChanHar-Peled2009}, since all there algorithms
have running-times of the form $n^{\poly(1/\eps)}$.
Furthermore, our algorithm easily extends to intersection
graphs of pseudo-disks, which is the class of graphs considered
in~\cite{ChanHar-Peled2009}.

The novelty behind the algorithm in Theorem~\ref{Theorem: Non-Crossing
Rectangles} lies in its usage of the \emph{arrangement
graph}~\cite{AgarwalSharir1998} of the input set of rectangles
$\calR$.  This graph is defined by considering all intersection points
occurring on boundary of rectangles as vertices, and the boundary
curves connecting them as edges.  By its definition, the arrangement
graph of a rectangle family is planar and $4$-regular, and thus has a
very convenient structure.  However, there is no immediate way to
translate approximate vertex-covers in the arrangement graph $A_\calR$
of $\calR$, to vertex covers in the corresponding rectangle graph
$G_\calR$.  Nevertheless, we show that we can translate
tree-decompositions in $A_\calR$ to tree-decompositions in $G_\calR$
of roughly the same width, and this allows with some technical effort
to simulate Baker's algorithm~\cite{Baker1994}. We believe that the
arrangement graph can be a useful tool in other intersection-graph
problems.

The second algorithm we present in this paper applies to general
rectangle families, and can handle also weights.  This algorithm
exploits the observation that the rectangles of a triangle-free
rectangle graph can be partitioned into two classes, where no pair of
rectangles cross in each class.  This, in combination with
Theorem~\ref{Theorem: Non-Crossing Rectangles} and the fact that we
can clean all triangles from our input graph at cost of a $1.5$ factor
to the approximation guarantee, gives us Theorem~\ref{Theorem: General
Rectangles} below for the unweighted case.  For the weighted case, we
use the additional observation that triangle-free non-crossing
rectangle graphs are planar, and so we can use Baker's
algorithm~\cite{Baker1994} directly.

\begin{theorem}
\label{Theorem: General Rectangles}%
Given any $\eps > 0$, \rvc\ can be approximated within a factor
of $1.5 + \eps$ in $2^{\poly(1/\eps)} \cdot \poly(n)$ time.
\end{theorem}


\section{Preliminaries}
\label{Section: Preliminaries}

We denote our input set of axis-parallel rectangles in the plane by
$\calR = \{R_1,\ldots,R_n\}$. We assume that each rectangle $R$ is
specified by two intervals $R=(X,Y)$, where $X$ is the projection of
$R$ on the $x$-axis, and $Y$ is the projection of $R$ on the
$y$-axis. We assume w.l.o.g. that $\calR$ is in \emph{general
position}, \emph{i.e.} that all intervals in the specification of
$\calR$ have different endpoints. The \emph{boundary} of a rectangle
$R$ is the set of all points with minimum and maximum $x$-coordinate
values, and minimum and maximum $y$-coordinate values.

\begin{figure}[h]
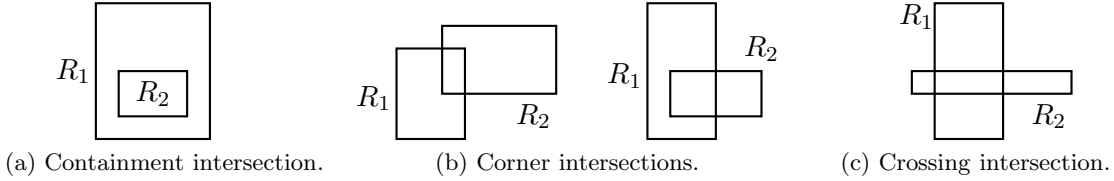

\begin{center}
\psset{unit=0.6}
\begin{tabular}{c@{\hspace{10pt}}c@{\hspace{10pt}}c}
\subfloat[Containment intersection.]{%
\pspicture(-2,0)(5,3)
  \psset{linestyle=solid,linecolor=black,arrows=-}
  \pspolygon(0,0)(0,3)(2.5,3)(2.5,0)
  \pspolygon(0.5,0.5)(0.5,1.5)(2,1.5)(2,0.5)
  \rput(-0.5,1.5){$R_1$}
  \rput(1.25,1){$R_2$}
\endpspicture
}
&
\subfloat[Corner intersections.]{%
\pspicture(-1,0)(8.5,3)
  \psset{linestyle=solid,linecolor=black,arrows=-}
  \pspolygon(0,0)(0,2)(1.5,2)(1.5,0)
  \pspolygon(1,1)(1,2.5)(3.5,2.5)(3.5,1)
  \rput(-0.5,0.9){$R_1$}
  \rput(3,0.5){$R_2$}
  \pspolygon(5.5,0)(5.5,3)(7,3)(7,0)
  \pspolygon(6,0.5)(6,1.5)(8,1.5)(8,0.5)
  \rput(5,1.4){$R_1$}
  \rput(8,2){$R_2$}
\endpspicture
}
&
\subfloat[Crossing intersection.]{%
\label{Fig:Rec4}
\pspicture(-2,0)(5,3)
  \psset{linestyle=solid,linecolor=black,arrows=-}
  \pspolygon(0.5,0)(0.5,3)(2,3)(2,0)
  \pspolygon(0,1)(3.5,1)(3.5,1.5)(0,1.5)
  \rput(3,0.5){$R_2$}
  \rput(0.1,2.7){$R_1$}
\endpspicture
}
\end{tabular}
\end{center}
\caption{Possible intersections between two rectangles $R_1$ and $R_2$.}
\label{Fig:Rec}
\end{figure}

Two rectangles $R_1 = (X_1,Y_1)$ and $R_2 = (X_2,Y_2)$
\emph{intersect}, denoted $R_1 \cap R_2 \neq \emptyset$, if they share
a common point, \emph{i.e.} if $X_1 \cap X_2 \neq \emptyset$ and $Y_1
\cap Y_2 \neq \emptyset$. Two non-intersecting rectangles are said to
be \emph{disjoint}.  There are three possible types of intersections
between two rectangles $R_1$ and $R_2$:
\begin{enumerate}
\item \emph{Containment intersection}: $R_1$ contains $R_2$. 
      In this case $R_1$ contains all corners of $R_2$, and the
      boundaries of $R_1$ and $R_2$ do not intersect.
\item \emph{Corner intersection}: $R_1$ contains one or two
    corners of $R_2$. In this case the boundaries of $R_1$ and $R_2$
    intersect exactly twice.
\item \emph{Crossing intersection}: the intersection of
    $R_1$ and $R_2$ does not involve any corners.  In this case, the
    boundaries of $R_1$ and $R_2$ intersect four times.
\end{enumerate}
See examples in Figure~\ref{Fig:Rec}.

Given a graph $G$, we use $V(G)$ and $E(G)$ to denote its vertex and
edge set, respectively. For a given vertex-subset $V \subseteq V(G)$,
we let $G[V]$ denote the subgraph of $G$ \emph{induced} by $V$,
\emph{i.e.} the subgraph with vertex-set $V$ and edge-set $\{\{u,v\}
\in E(G) : u,v \in V\}$.  We write $G-V$ to denote the induced
subgraph $G[V(G) \setminus V]$.  We will also be considering
vertex-weighted graphs, \emph{i.e.}  graphs $G$ equipped with a
weight-function $w: V(G) \to \mathbb{Q}$.  A \emph{vertex cover} of
$G$ is a subset of vertices $C \subseteq V(G)$ such that $\{u,v\} \cap
C \neq \emptyset$ for any edge $\{u,v\} \in E(G)$. For a non-negative
real $\alpha \in \mathbb{R}^{\geq 0}$, an \emph{$\alpha$-approximate
vertex cover} of $G$ is a vertex cover $C$ with $|C| \leq \alpha \cdot
\opt$ (or $w(C) \leq \alpha \cdot \opt$ in the weighted case), where
$\opt$ is the size (weight) of an optimal vertex cover of $G$.  The
\emph{intersection graph} $G_\calR$ corresponding to our input set of
rectangles $\calR$ is the graph with vertex-set $V(G_\calR)=\calR$,
and edge-set $E(G_\calR)=\{\{R_1,R_2\}: R_1 \cap R_2 \neq \emptyset
\}$.

We will be using an important tool due to Nemhauser and
Trotter~\cite{NemhauserTrotter1975} that allows us to focus on
graphs whose entire vertex-set already constitutes a good
approximate vertex-cover:


\begin{theorem}[Nemhauser$\&$Trotter~\cite{NemhauserTrotter1975}]
\label{Theorem: NT}%
There is a polynomial-time algorithm that given a (vertex-weighted)
graph $G$, computes a vertex set $V \subseteq V(G)$ such that:
\begin{enumerate}[(i)]
\item $V$ is a 2-approximate vertex-cover of $G[V]$, and
\item any $\alpha$-approximate vertex cover of $G[V]$ can be converted 
      in polynomial-time to an $\alpha$-approximate vertex cover
      of $G$.
\end{enumerate}
\end{theorem}

Finally, we will be using the notion of treewidth and
tree-decomposition of graphs, introduced in the form below by
Robertson and Seymour~\cite{RobertsonSeymour1986}.

\begin{definition}[Tree Decomposition, Treewidth~\cite{RobertsonSeymour1986}]
\label{Definition: Treewidth}%
A \emph{tree decomposition} of a graph $G$ is a pair $(\calT,\calX)$,
where $\calX \subseteq 2^{V(G)}$ is family of vertex subsets of $G$,
and $\calT$ is a tree over $\calX$, satisfying the following
conditions:
\begin{enumerate}
\item $\bigcup_{X \in \calX} G[X] = G$, and
\item $\calX_v = \{X \in \calX : v \in X\}$ is connected in $\calT$ 
      for all $v \in V(G)$.
\end{enumerate}
The \emph{width} of $\calT$ is $\max_{X \in \calX}|X|-1$.  The
\emph{treewidth} of $G$, denoted $\tw(G)$, is the minimum width over
all tree decompositions of $G$.
\end{definition}


\section{An EPTAS for Non-Crossing Rectangle Graphs}
\label{Section: EPTAS}

In this section we present an EPTAS for \rvc\ in unweighted
non-crossing rectangle families. 
This algorithm extends to intersection graphs of pseudo-disks.

Let $\calR$ denote our input set of unweighted non-crossing
rectangles.  The first step of our algorithm is to clean $\calR$ from
containment intersections and pairwise intersecting subsets of size
greater than some constant $q \geq 2$ to be chosen later. This can be
done using standard techniques, and allows us to gain substantial
structure at a small cost to the approximation factor of our
algorithm.

\begin{lemma}
\label{Lemma: Cleaning Step}%
Suppose that \rvc\ in corner-intersecting rectangle families with no
$q+1$ pairwise intersecting rectangles can be approximated within a
factor of $\alpha$. Then \rvc\ in non-crossing rectangle families can
be approximated within a factor of $\max \{\alpha,1+1/q\}$ in
polynomial-time.
\end{lemma}

\begin{proof}
Let $\calR$ be any non-crossing rectangle family. 
First observe that if $\calR$ has two rectangles $R_1$ and $R_2$ such
that $R_1$ is contained in $R_2$, then (in the unweighted case) we may
assume w.l.o.g.\ that any vertex cover of $G_\calR$ includes $R_2$.
Indeed, if $\calC$ is a vertex-cover of $G_\calR$ that does not
contain $R_2$, then $R_1 \in \calC$, and therefore $\calC'=(\calC
\setminus \{R_1\})\cup \{R_2\}$ is also a vertex-cover of $G_\calR$.
%
Second, note that if $\calQ_1,\calQ_2,\ldots,\calQ_r$ are cliques of
size $q+1$ in $G_\calR$ such that $\calQ_i \cap \calQ_j = \emptyset$
for every $i \neq j$, then any vertex-cover of $G_\calR$ must include
at least $r \cdot q$ rectangles from $\bigcup_{1 \leq i
\leq r} \calQ_i$.  Thus, by including all rectangles from these
cliques in our solution, we deviate (again, in the unweighted
case) by at most a factor of $1+1/q$ from the optimum.

Using these two observations, we proceed as follows: We compute two
sets of rectangles $\calP,\calQ \subseteq \calR$.  The set $\calP$
includes all rectangles of $\calR$ that contain other rectangles of
$\calR$, and $\calQ$ is a greedy packing of disjoint $(q+1)$-cliques
in $G_{\calR \setminus \calP}$. Observe that both these sets can be
computed in polynomial-time, $\calP$ by obvious brute force, and
$\calQ$ by the fact that every clique in $\calR$ is represented by
some intersection point of two rectangles (or even by a rectangle
corner as $\calR \setminus \calP$ has only corner intersections).  We
then apply the $\alpha$-approximation algorithm assumed in the lemma
to obtain a $\alpha$-approximate vertex-cover $\calC$ for $\calR
\setminus (\calP \cup \calQ)$.  According to the two observations
above, $\calP \cup \calQ \cup \calC$ is a $\max
\{\alpha,1+1/q\}$-approximate vertex-cover of $G_\calR$.
\end{proof}

Due to Lemma~\ref{Lemma: Cleaning Step} above, we can henceforth
assume that $\calR$ contains only corner intersections, and that the
maximum clique in $G_\calR$ is of size at most $q$. We also apply the
Nemhauser\&Trotter algorithm (Theorem~\ref{Theorem: NT}) on $G_\calR$
after applying Lemma~\ref{Lemma: Cleaning Step}, and so we assume that
$\calR$ is a 2-approximate vertex cover of $G_\calR$.

The main idea of algorithm is as follows. We will construct the
so-called \emph{arrangement graph} $A_\calR$ of $\calR$ which is build
by considering all intersection points occurring on boundaries of
rectangles as vertices, and the boundary curves connecting them as
edges.  By this construction, $A_\calR$ is a planar graph, and as
such, it has very specific structure. The most tempting approach is to
use Baker's EPTAS for \vc\ in planar graphs on $A_\calR$, and to
convert the $(1+\eps)$-approximate vertex-cover of $A_\calR$ to a
$(1+\eps')$-approximate vertex-cover of $G_\calR$.  Unfortunately,
this attempt fails, since the natural transformation from vertices of
$A_\calR$ to rectangles of $G_\calR$ produces the entire set of
rectangles $\calR$ on any vertex-cover of $A_\calR$. We therefore take
an alternative route.  The basic idea is to mimic Baker's algorithm by
using the observation that tree-decompositions of $A_\calR$ correspond
to tree-decompositions of $G_\calR$ of roughly the same width.  Thus,
instead of applying Baker's algorithm on $A_\calR$ as a black-box, we
can simulate its steps directly on $G_\calR$.  Using then an extension
of Baker's analysis, we can show that this approach indeed gives us
our the desired $(1+\eps)$ approximation factor.


\subsection{The arrangement graph}
\label{Subsection: arrangement graph}

\begin{figure}[h]
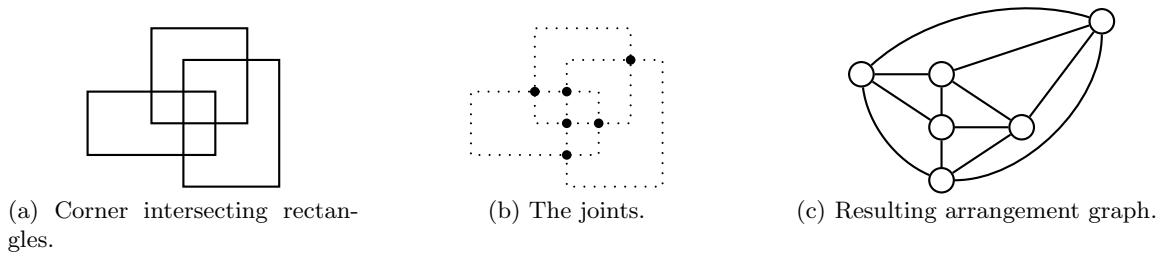

\begin{center}
\psset{unit=0.42}
\begin{tabular}{ccc}
\subfloat[Corner intersecting rectangles.]{%
\pspicture(-2.5,0)(8.5,5)
  \psset{linestyle=solid,linecolor=black,arrows=-}
  \pspolygon(3,0)(6,0)(6,4)(3,4)
  \pspolygon(0,1)(4,1)(4,3)(0,3)
  \pspolygon(2,2)(5,2)(5,5)(2,5)
\endpspicture
}
&
\subfloat[The joints.]{%
\pspicture(-2.5,0)(8.5,5)
  \psset{linestyle=dotted,linecolor=black,arrows=-}
  \pspolygon(3,0)(6,0)(6,4)(3,4)
  \pspolygon(0,1)(4,1)(4,3)(0,3)
  \pspolygon(2,2)(5,2)(5,5)(2,5)
  \psdot(2,3)
  \psdot(3,1)
  \psdot(3,2)
  \psdot(3,3)
  \psdot(4,2)
  \psdot(5,4)
\endpspicture
}
&
\subfloat[Resulting arrangement graph.]{
\pspicture(-2,-0.2)(2,6)
\psmatrix[mnode=C,radius=5pt,colsep=20pt,rowsep=10pt]
  &   &   & 1 \\
2 & 3 \\
  & 4 & 5 \\
  & 6
\endpsmatrix
\psset{shortput=nab,arrows=-}
\ncarc[arcangle=-30]{1,4}{2,1}
\ncline{1,4}{2,2}
\ncline{1,4}{3,3}
\ncarc[arcangle=45]{1,4}{4,2}
\ncline{2,1}{2,2}
\ncline{2,1}{3,2}
\ncarc[arcangle=-30]{2,1}{4,2}
\ncline{2,2}{3,2}
\ncline{2,2}{3,3}
\ncline{3,2}{3,3}
\ncline{3,2}{4,2}
\ncline{3,3}{4,2}
\endpspicture
}
\end{tabular}
\end{center}
\caption{A set $\calR$ of corner intersection rectangles and its
arrangement graph $A_\calR$.}
\label{Fig:arrangement}
\end{figure}

In this section we present several properties of \emph{arrangement
graphs}~\cite{AgarwalSharir1998} of rectangle families.
An intersection of two rectangle boundaries is called a \emph{joint}.
The \emph{arrangement graph} $A_\calR$ of a rectangle family $\calR$
is the graph that is defined as follows: The vertex set of $A_\calR$
is the set of joints.  The edge set of $A_\calR$ consists of the
rectangle boundary fragments, namely $\{u,v\}$ is an edge in $A_\calR$
if and only if $u$ and $v$ are two joints located on the boundary of
some rectangle such that no other joint is located on the boundary
between them. It is not difficult to see that the arrangement graph
defined as above is in fact planar and $4$-regular (see example in
Figure~\ref{Fig:arrangement}).

For a given subset of joints $J \subseteq V(A_\calR)$, the set of
rectangles that is induced by $J$ is defined by 
\[
\calR(J) = 
\{R \in \calR ~:~ \exists j \in J \text{ s.t. } j 
                  \text{ is on the boundary of
} R\}
~. 
\]
The following lemma is immediate from the fact that $\calR$ is
in general position.

\begin{lemma}
\label{Lemma: Induced Rectangles}%
$|\calR(J)|=2 \cdot |J|$ for any set of joints $J \subseteq
V(A_\calR)$.
\end{lemma}

The following lemma states that the number of joints is $A_\calR$ is
linear in $|\calR|$.

\begin{lemma}
\label{Lemma: Arrangement Graph Size}%
$|V(A_\calR)| \leq 4q \cdot |\calR|$.
\end{lemma}

\begin{proof}
Recall that we assume $\calR$ to have only corner intersections and no
$q+1$ pairwise intersecting rectangles. Now, if two rectangles $R_1$
and $R_2$ are corner intersecting, then the boundaries of $R_1$ and
$R_2$ intersect exactly twice. Hence, $|V(A_\calR)| = 2|E(G_\calR)|$.
Furthermore, since each corner can be involved in at most $q$
intersections, we have $\sum_{R \in \mathcal{R}} \deg(R) \leq 4q$,
where $\deg(R)$ denotes the number of rectangles intersecting $R$.
Thus, $|E(G_\calR)| \leq 2q \cdot |\calR|$, and so $|V(A_\calR)| \leq
2 \cdot |E(G_\calR)| \leq 4q \cdot |\calR|$.
\end{proof}


\subsection{Baker's Algorithm}
\label{Subsection: The Algorithm}

Our algorithm for \rvc\ in non-crossing rectangle families
simulates Baker's classical algorithm for \vc\ in planar
graphs~\cite{Baker1994} on the arrangement graph $A_\calR$ of
$\calR$. The main idea behind Baker's approach is the
observation that given a planar graph $G$ and any positive
integer $k$, one can partition the vertex-set of $G$ into $k$
classes such that deleting each the vertices in class results
in subgraph of treewidth at most $3k$ (see Lemma~\ref{Lemma:
Baker Partitioning} below). Combining this observation along
with the well-known algorithm for \vc\ in bounded treewidth
graphs (see \emph{e.g.}~\cite{Niedermeier2006}), gives an EPTAS
for \vc\ in planar graphs.

\begin{lemma}[Baker~\cite{Baker1994}]
\label{Lemma: Baker Partitioning}%
Given a planar graph $G$ and an integer $k$, one can partition $V(G)$
into $k$ subsets $V_1,\ldots,V_k$ such that $\tw(G-V_i) \leq 3k$ in
polynomial-time with respect to both $|V(G)|$ and $k$.
\end{lemma}

In order to properly simulate Baker's approach on \rvc\, we will need
a slightly more general framework. In particular, our algorithm will
not necessarily produce a partition of the vertex-set of
$G_\calR$. Also, our algorithm will produce vertex-sets whose deletion
results in a subgraph of $G_\calR$ with treewidth slightly more than
$3k$. Nevertheless, it is not difficult to show that a slight
relaxation of these two requirements does not alter Baker's analysis
too much:

\begin{lemma}
\label{Lemma: Baker Extension}%
Let $G$ be a graph with $n$ vertices, and let $c_1$ and $c_2$ be two
fixed positive integers. Suppose that there is a polynomial-time
algorithm that, given $G$ and a positive integer $k$, produces
vertex-sets $U_1,\ldots,U_k$ with the following properties:
\begin{enumerate}
\item $\bigcup_i U_i = V(G)$.
\item $\sum_i |U_i| \leq c_1 \cdot n$.
\item $\tw(G-U_i) \leq c_2 \cdot k$ for every $i$.
\end{enumerate}
Then one can compute an vertex-cover of $G$ within a factor of
$(1+\eps)$ in $2^{\poly(1/\varepsilon)} \cdot \poly(n)$ time, for any
given $\eps > 0$.
\end{lemma}

\begin{proof}
Choose $k$ to be an integer greater or equal to $2c_1/\eps$, and let
$U_1,\ldots,U_k$ be the vertex-sets produced by the algorithm assumed
in the lemma.  We assume $G$ has been preprocessed using the
Nemhauser\&Trotter algorithm (Theorem~\ref{Theorem: NT}), and thus $n
\leq 2\opt$, where $\opt$ is the size of the minimum vertex-cover of
$G$. For each $i$, $1 \leq i \leq k$, let $H_i$ denote the subgraph
$H_i=G_i - U_i$, and let $\opt_i$ be the size of the minimum
vertex-cover of $H_i$.  Since $\tw(H_i) \leq c_2 \cdot k$, a
vertex-cover $C^*_i$ for $H_i$ of size $\opt_i$ can be computed in
$2^{\poly(1/\eps)} \cdot \poly(n)$ time.  We thus obtain $k$ candidate
vertex-covers for $G$, $C_i = U_i \cup C^*_i$, and we have
\[
\sum_i |C_i| 
=    \sum_i (|U_i| + \opt_i) \leq c_1 n + \sum_i \opt_i
\leq 2 c_1  \opt + k \opt 
=    (k+2c_1)\opt.
\]
Therefore, choosing the smallest among the $C_i$'s, we get a
vertex-cover for $G$ of size at most $\min_i |C_i| \leq (1+2c_1/k)
\opt \leq (1+\eps) \opt$.
\end{proof}


\subsection{Our Algorithm}
\label{Subsection: The Analysis}

We are now in position to describe our EPTAS.  The key lemma we
need is Lemma~\ref{Lemma: Treewidth} below that allows us to convert
tree-decompositions of $A_\calR$ to a tree-decompositions of $G_\calR$
of approximately the same width.

\begin{lemma}
\label{Lemma: Treewidth}%
$\tw(G_\calR) \leq 2 \cdot \tw(A_\calR) + 1$.
\end{lemma}
\begin{proof}
Let $(\calT,\calX)$ is a tree-decomposition of $A_\calR$ whose width
it $\tw(A_\calR)$.
Now let $\calX' = \{\calR(X): X \in \calX\}$, and let $\calT'$ be a
tree over $\calX'$ with an edge $\{\calR(X_1),\calR(X_2)\}$ for every
edge $\{X_1,X_2\}$ in $\calT$.  We show that $(\calT',\calX')$ is a
tree-decomposition of $G_\calR$, namely that $(\calT',\calX')$
satisfies all requirements of Definition~\ref{Definition:
Treewidth}. First, observe that any rectangle has at least two
corresponding joints since we assume there are no isolated rectangles
in $\calR$. Furthermore, if two rectangles intersect, then there is a
joint $j \in V(A_\calR)$ that corresponds to both these
rectangles. Hence, for every edge $\{R_1,R_2\} \in E(G_\calR)$, there
is at least one node in $\calX'$ which contains both $R_1$ and $R_2$.
Thus, $\bigcup_{X \in \calX'} G_\calR[X]= G_\calR$.

Now suppose there is some rectangle $R$ which is contained in
two nodes $\calR(X_1)$ and $\calR(X_2)$ of $\calT'$. Then $R$
has two joints $j_1$ and $j_2$ with $j_1 \in X_1$ and $j_2 \in
X_2$. By construction, there is a path $j_1,i_1,\ldots,i_r,j_2$
connecting $j_1$ to $j_2$ in $A_\calR$, where $i_1,\ldots,i_r$
are all joints of $R$. Since $(\calT,\calX)$ is a proper tree
decomposition of $A_\calR$, it follows that there is a path
$X_1,Y_1,\ldots,Y_{r'},X_2$ connecting $X_1$ and $X_2$ in
$\calT$, with $Y_i \cap \{j_1,i_1,\ldots,i_r,j_2\} \neq
\emptyset$ for each $i$, $1 \leq i \leq s$. Thus, each node in
the path $\calR(X_1), \calR(Y_1), \ldots, \calR(Y_{s}),
\calR(X_2)$ connecting $\calR(X_1)$ and $\calR(X_2)$ in
$\calT'$ contains $R$, and since $R$, $\calR(X_1)$, and
$\calR(X_2)$ were chosen arbitrarily, this shows that for each
$R \in \calR$:
\[
\{ \calR(X) \in \calX' : R \in X\} \text{ is connected in } \calT'.
\]
Thus, both requirements of Definition~\ref{Definition:
Treewidth} are fulfilled by $(\calT',\calX')$.
 
Finally, observe that due to Lemma~\ref{Lemma: Induced Rectangles},
$\max_{X' \in \calX'} |X'| \leq 2\max_{X \in \calX} |X|$.  It follows
that the width of $(\calT',\calX')$ is at most $2\tw(A_\calR) + 1$,
and so the lemma is proved.
\end{proof}

Our algorithm consists of the following steps:
\begin{enumerate}
\item Set $q = \ceil{1/\eps}$ and $k = \ceil{8q/\eps} = \ceil{8/\eps^2}$.
\item Apply Lemma~\ref{Lemma: Cleaning Step} so that
      $\calR$ does not have any containment intersections and no
      pairwise intersecting subsets of rectangles of size greater than
      $q$.
\item Apply the Nemhauser\&Trotter Theorem on $G_\calR$,
      and let $\calR' \subseteq \calR$ denote the resulting subset of
      rectangles.
\item Construct the arrangement graph $A_{\calR'}$ corresponding to 
      $\calR'$, and partition $A_{\calR'}$ into $k$ subsets
      $V_1,\ldots,V_k$ using Lemma~\ref{Lemma: Baker Partitioning}.
\item Use Lemma~\ref{Lemma: Baker Extension} on $G_{\calR'}$
      with $U_i = \calR'(V_i)$, for every $i$.
\end{enumerate}

Observe that the arrangement graph of $\calR' \setminus U_i$ is a
subgraph of $A_{\calR'} - V_i$.  Hence, according to
Lemmas~\ref{Lemma: Induced Rectangles}, \ref{Lemma: Arrangement Graph
Size} and~\ref{Lemma: Treewidth} above, $U_1,\ldots,U_k$ satisfy the
three conditions of Lemma~\ref{Lemma: Baker Extension}.  Thus the
above algorithm outputs a $(1+\eps)$-approximate vertex-cover of
$G_\calR$ in $2^{\poly(1/\eps)} \cdot \poly(n)$ time.  This proves
Theorem~\ref{Theorem: Non-Crossing Rectangles}.

Finally, we mention that our EPTAS can be modified to deal with
intersection graphs of pseudo-disks.  Specifically, in Step~2, instead
of removing cliques, we remove point cliques, namely subsets of
rectangles $\calQ$ such that $\bigcap_{R \in \calQ} R \neq \emptyset$.
This is sufficient, since the number of joints in the arrangement
graph $A_\calD$ of a set $\calD$ of pseudo-disks, where no point is
contained in more than $q$ pseudo-disks, is $O(q \cdot
|\calD|)$~\cite{SharirAgarwal95}.


\section{General Rectangle Graphs}
\label{Section: General Rectangle Graphs}

In this section we present an algorithm for \rvc\ in general
rectangle families.  Our algorithm achieves an approximation
factor of $1.5 + \eps$, for any given $\varepsilon >0$, in time
$2^{\poly(1/\eps)} \cdot \poly(n)$, and works also for the
weighted variant of the problem. 

We begin with the unweighted case, and with the following lemma which
relies on an observation already made by Lewin-Eytan \emph{et
al.}~\cite{Lewin-EytanNaorOrda2002}.  A rectangle family is said to be
\emph{triangle-free} if there are no three pairwise intersecting
rectangles in the family.

\begin{lemma}
\label{Lemma: Non-crossing Partition}%
Any triangle-free rectangle family can be partitioned into two
non-crossing subsets in polynomial time.
\end{lemma}
\begin{proof}
Let $\calR$ be a triangle-free rectangle family. Observe that any two
rectangles $R_1 = (X_1,Y_1)$ and $R_2 = (X_2,Y_2)$ are crossing if and
only if either
\begin{itemize}
\item $X_1 \subset X_2$ and $Y_2 \subset Y_1$, or
\item $X_2 \subset X_1$ and $Y_1 \subset Y_2$,
\end{itemize}
It follows that the crossing relation between rectangles forms a
partial order. Thus, by Dilworth's Theorem~\cite{Dilworth1950}, and
due to the fact $\calR$ is triangle-free, there exists a partitioning
$\Pi$ of $\calR$ into two non-crossing subsets as desired.
Furthermore, using one of many classical minimum anti-chain
partitioning algorithms (see \emph{e.g.}~\cite{Golumbic1980}), one can
compute $\Pi$ in polynomial-time. 
\end{proof}

Observe that Lemma~\ref{Lemma: Non-crossing Partition} is already
enough, along with the results in Section~\ref{Section: EPTAS}, to
obtain our desired $1.5 + \eps$ approximation factor.  The algorithm
proceeds in the following six steps, given a rectangle family $\calR$
and $\eps >0$:
\begin{enumerate}
\item Apply Lemma~\ref{Lemma: Cleaning Step} to obtain a 
      triangle-free rectangle family $\calR' \subseteq \calR$.

\item Apply the Nemhauser\&Trotter algorithm to obtain a subset 
      $\calR'' \subseteq \calR'$ which is a 2-approximate vertex-cover
      of $G_{\calR''}$.

\item Use Lemma~\ref{Lemma: Non-crossing Partition} to obtain a 
      partitioning $\{\calR_1,\calR_2\}$ of $\calR''$, where both
      $\calR_1$ and $\calR_2$ are non-crossing.

\item Compute an $\eps$-approximate vertex-cover $\calC_1$ of 
      $G_{\calR_1}$ and an $\eps$-approximate vertex-cover $\calC_2$
      of $G_{\calR_2}$ using the EPTAS of Section~\ref{Section:
      EPTAS}.

\item Use the best of the two vertex-covers $\calR_1 \cup \calC_2$ and 
      $\calR_2 \cup \calC_1$ for $G_{\calR''}$, along with the
      Nemhauser\&Trotter Theorem to compute a vertex-cover of
      $G_{\calR'}$.

\item Add the removed rectangles as required by 
      Lemma~\ref{Lemma: Cleaning Step} to obtain a vertex-cover for
      $G_\calR$.
\end{enumerate}

The fact that this algorithm outputs a vertex-cover which is a factor
of $1.5 + \eps$ off the optimum follows from a similar analysis used
in Lemma~\ref{Lemma: Baker Extension}.  Clearly, both $\calR_1 \cup
C_2$ and $\calR_2 \cup C_1$ are vertex covers for $G_{\calR''}$.
Furthermore, letting $\opt$, $\opt_1$, and $\opt_2$ denote the size of
the minimum vertex-covers of $G_{\calR''}$, $G_{\calR_1}$, and
$G_{\calR_2}$ respectively, we get:
\begin{align*}
|\calR_1 \cup C_2| + |\calR_2 \cup C_1| 
& \leq |\calR| + (1+\eps) \opt_1 + (1+\eps) \opt_2 \\ 
& \leq 2\opt + (1+\eps)\opt \\ 
& =    (3+\eps)\opt
~.
\end{align*}
Thus, the minimum of both $\calR_1 \cup \calC_2$ and $\calR_2 \cup
\calC_1$ gives a $(1.5 + \eps)$-approximate vertex-cover for
$G_{\calR''}$.  Applying the Nemhauser\&Trotter Theorem along with
Lemma~\ref{Lemma: Cleaning Step} shows that the algorithm above
outputs a $(1.5 + \eps)$-approximate vertex-cover for $G_\calR$.

For the weighted variant of the problem, we observe that all steps of
the algorithm above, apart from Step~4, can be applied also in the
weighted case.  For the first step we use a weighted version of
Lemma~\ref{Lemma: Cleaning Step}, which can be obtained by a standard
application of the local-ratio technique~\cite{Bar-YehudaEven1985}
(see \emph{e.g.}~\cite{ButmanHermelinLewensteinRawitz2007}).  All
other steps have immediate weighted counterparts.  To replace Step~4,
we use the following observation that the intersection graph of any
triangle-free non-crossing rectangle family is planar:

\begin{lemma}
\label{Lemma: Planar}%
If $\calR$ is triangle-free and non-crossing, then $G_\calR$ is
planar.
\end{lemma}

\begin{proof}
Notice that since $\calR$ is triangle-free, $R_i \cap R_j \cap R_k =
\emptyset$ for every three rectangles $R_i,R_j,R_k \in \calR$.  In
other words, every point $p$ in the plane is contained in at most two
rectangles.  Moreover, observe that $R_i \cap R_j$ is a rectangle for
every two intersecting rectangles $R_i$ and $R_j$ (see
Figure~\ref{Fig:Rec}).

Assume, without loss of generality, that the rectangles in $\calR = \{
R_1,\ldots,R_n \}$ are numbered in such a way that if $R_i \subseteq
R_j$, then $i < j$.  We construct a family $\calR = \{
R'_1,\ldots,R'_n \}$ of rectilinear polygons as follows.  For $i < j$
let $R_{ij}$ be the set of points in $R_i \cap R_j$ that are not on
the border of $R_i$.  We define $R'_j = R_j \setminus \cup_{i<j}
R_{ij}$.  Notice that if $R_i$ is contained in $R_j$, then $R_i' =
R_i$.  Hence, $R_i' \neq \emptyset$ for every $i$.  Furthermore,
observe that $R_i$ and $R_j$ intersect if and only if $R'_i$ and
$R'_j$ intersect.  Hence, $G_{\calR'} \cong G_\calR$.  Finally, notice
that intersections in $\calR'$ only contain polygon borders. It
follows that $G_{\calR'}$ is the dual of a planer graph, and thus
itself is planar, and therefore $G_\calR$ is also planar.
\end{proof}

Thus, according to Lemma~\ref{Lemma: Planar} above, we can apply
Baker's algorithm for \vc\ in planar graphs instead of our EPTAS in
step 4 of the algorithm above.  Indeed, Baker's algorithm can also
handle weights.  Thus, by the same analysis given above, we get a $1.5
+ \eps$ for the weighted variant of $\rvc$. We mention also that
Lemma~\ref{Lemma: Planar} above can also be used to obtain a
1.5-approximation algorithm for the weighted variant of $\rvc$ in
non-crossing rectangle families.

\section*{Acknowledgments}
We thank Micha Sharir for helpful discussions.


\bibliographystyle{abbrv}

\begin{small}

\end{small}

\end{document}